\documentclass[twocolumn,english,aps,nofootinbib]{revtex4}
\usepackage[T1]{fontenc}
\usepackage[latin9]{inputenc}
\usepackage{amsmath}
\usepackage{amssymb}
\usepackage{appendix}
\usepackage{esint}
\usepackage{tikz}
\usepackage{graphicx}
\usepackage{relsize}

\makeatletter

\@ifundefined{textcolor}{}
{%
 \definecolor{BLACK}{gray}{0}
 \definecolor{WHITE}{gray}{1}
 \definecolor{RED}{rgb}{1,0,0}
 \definecolor{GREEN}{rgb}{0,1,0}
 \definecolor{BLUE}{rgb}{0,0,1}
 \definecolor{CYAN}{cmyk}{1,0,0,0}
 \definecolor{MAGENTA}{cmyk}{0,1,0,0}
 \definecolor{YELLOW}{cmyk}{0,0,1,0}
 }

\usepackage{nicefrac}\usepackage{dsfont}
\usepackage[normalem]{ulem}\@ifundefined{definecolor}
 {\usepackage{color}}{}

\makeatother

\providecommand{\keyword}[1]{\textbf{\textit{Keywords: }} #1}

\usepackage{babel}
\begin{document}


\title{Escape rate of transiently active Brownian particle in one dimension}
\author{A. Scacchi$^{1,2}$}
\author{J. M. Brader$^{1}$}
\author{A. Sharma$^{1,3}$}
\affiliation{1. Department of Physics, University of Fribourg, Fribourg, Switzerland\\2.Department of Mathematical Sciences, Loughborough University, Loughborough LE11 3TU, United Kingdom\\3. Leibniz-Institut f\"ur Polymerforschung Dresden, 01069 Dresden, Germany
}


\begin{abstract}
Activity significantly enhances the escape rate of a Brownian particle over a potential barrier. Whereas constant activity has been extensively studied in the past, little is known about the effect of time-dependent activity on the escape rate of the particle.  In this paper we study the escape problem for a Brownian particle that is transiently active; the activity decreases rapidly during the escape process. Using the effective equilibrium approach we analytically calculate the escape rate, under the assumption that the particle is either completely passive or fully active when crossing the barrier. We perform numerical simulations of the escape process in one dimension and find good agreement with the theoretical predictions. 
\end{abstract}

\maketitle

\keyword{Active Brownian particles, mean first passage time, effective equilibrium, escape process, time dependent activity}


\section{Introduction}
The escape rate of a Brownian particle over a potential barrier is accurately described by Kramers theory \cite{kramers1940brownian}. It states that for sufficiently large potential barriers and high viscosity, the escape rate decreases exponentially with the barrier height. The escape problem has been recently studied in the context of active Brownian particles (ABPs), which undergo self propulsion in addition to Brownian motion \cite{sharma2017escape, scacchi_sharma, caprini2019active}. Contrary to Brownian particles, ABPs are subject to both Brownian motion and a self-propulsion, which requires a continual consumption of energy from the surroundings~\cite{PhysRevLett.105.088304, Erbe_2008, PhysRevLett.99.048102, nature_1}. The direction of self propulsion is randomized on a finite time scale which is referred to as the persistence time. Due to self propulsion, active particles escape the potential barrier at a much higher rate than their passive counterparts. A quantitative description of the escape rate was provided in Ref. \cite{sharma2017escape} in which the authors derived a Kramers-like rate expression for the escape of an active particle in the limit of small persistence time, where the velocities are represented by a stochastic variable and the orientations are not considered explicitly. In the case of small activity the steady-state properties that this model has predicted revealed fascinating similarities with an equilibrium system~\cite{PhysRevLett.117.038103}. It is also worth mentioning that several sedimentation and trapping problems can be treated analytically on the single-particle level~\cite{PhysRevE.90.012111}. A more general theoretical description of the escape process, for both short and long persistence times, has been recently reported in Ref. \cite{caprini2019active}.

The escape of an ABP over a potential barrier has been investigated for a time independent activity~\cite{sharma2017escape, scacchi_sharma, caprini2019active}. However, it is perhaps more reasonable to assume that the particle loses activity over time and becomes passive. In a biological context, for example, the activity of a molecular motor inside animal cells depends on the supply of fuel molecules which may decrease with time \cite{molecular_motors}. Time dependent activity also features in the energy depot model \cite{schweitzer1998complex}. Furthermore, transient activity can be obtained in synthetic ABPs such as light driven Janus particles that can be made passive by simply switching off the laser \cite{lozano2016phototaxis}. Recently, time-varying activity fields have been applied to non-interacting ABPs in three-dimension, showing how travelling activity waves can induce fluxes \cite{varying_field}. Rapidly changing activity is observed in the translocation of a polymer chain across a membrane, on one side of which, molecular motors exert force on the polymer~\cite{PMID_poly}.

Motivated by these considerations, we study in this paper the escape problem for an active Brownian particle that rapidly becomes passive. That the activity is transient naturally introduces a time scale in the system: the time scale over which the particle can be considered active. If this time scale is larger than the escape time of the particle over the barrier, one can effectively ignore the transient nature of the activity. On the contrary, if the time scale of activity is very short, the particle escapes the barrier as a passive particle. Whereas the former corresponds to escape of an active particle over the potential barrier \cite{sharma2017escape, scacchi_sharma, caprini2019active}, the latter is the original Kramers escape problem. The most interesting scenario corresponds to the intermediate time scale of activity such that the escape process is partially active and partially passive.

In this paper, we focus on the escape problem in one dimension for a transiently active Brownian particle. We will show that a quantitative description of the escape rate can be obtained for the transiently active particle over a wide range of activity time scale. Assuming that the particle escapes over the barrier either as an active particle or as a passive particle, the escape rate can be calculated using the effective equilibrium approach \cite{sharma2017escape, farage2015effective}. This approach is a coarse grained approach in which the orientational degree of freedom of the particle is integrated out and has been successfully applied to the escape problem of active particles \cite{sharma2017escape}.

\section{Model and theory}\label{model}
\begin{figure}[h!]
\includegraphics[width=0.5\textwidth]{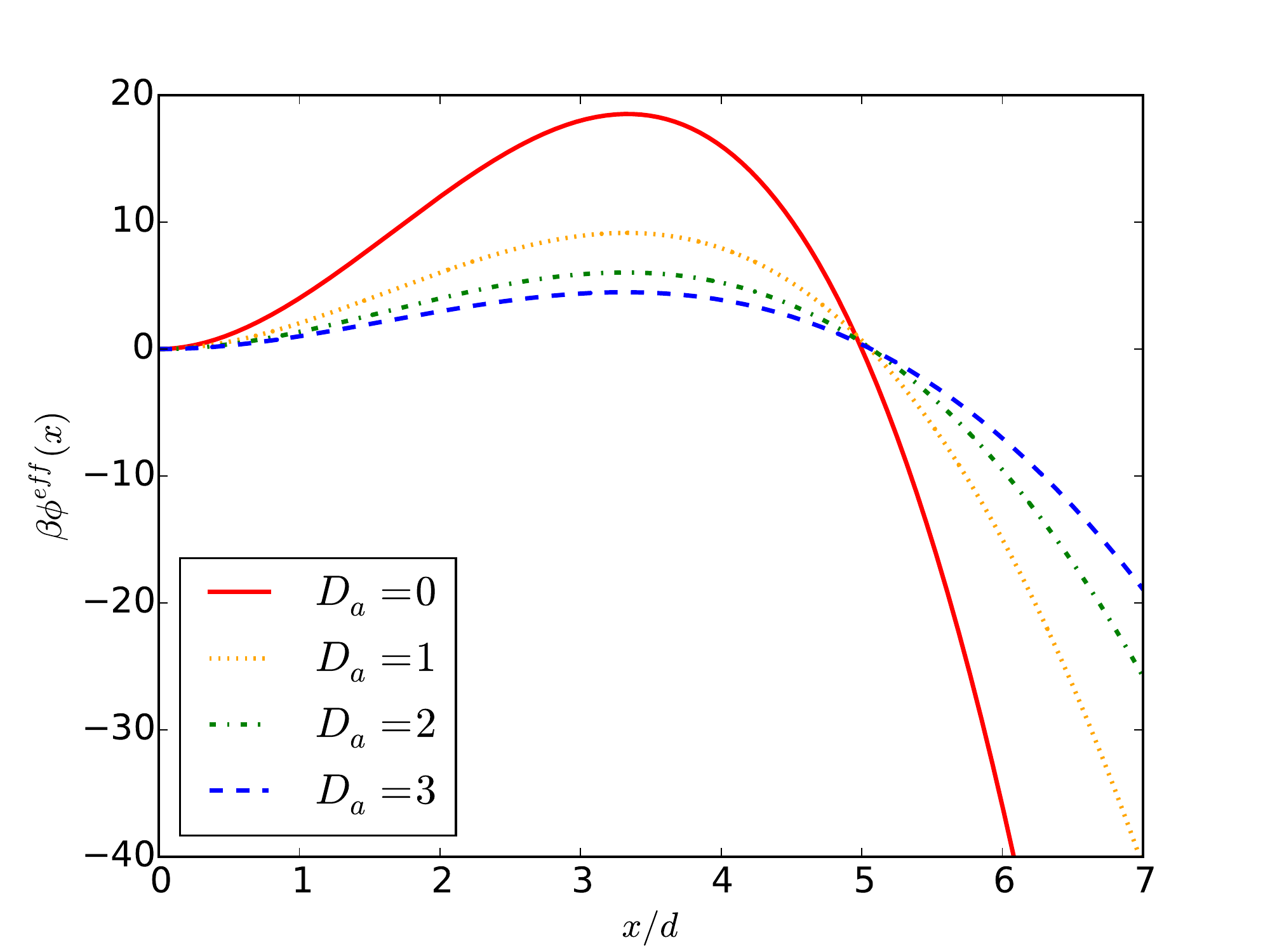}
\caption{Effective potentials $\phi^{eff}$ for different values of $D_a$ obtained from Eq.~(\ref{Vx}). The solid red curve ($D_a=0$) is a special case of Eq.~(\ref{Vx}) and corresponds to Eq.~(\ref{potential}). Here $\alpha=1$, $\omega_0=10$ and $\tau=0.01$. The results presented below are obtained for the value $\tau=0.05$ which is different from that for which the potential is sketched here. For $\tau = 0.05$, the effective potential exhibits an unphysical divergence at a location beyond the potential barrier. The divergence is, however, irrelevant for the performed analytics. We therefore choose $\tau=0.01$ in this figure to avoid confusion.}
\label{sketch}
\end{figure}

We consider a one dimensional system of a single Brownian particle with time dependent activity $v(t)$ with coordinate $x$ and orientation specified by an embedded unit vector $p$. The orientation vector can point either along the positive $x$-axis or the negative one. The switching between these two states occurs at an average rate of $\tau^{-1}$ in an independent, uncorrelated fashion. The motion of the particle can be modelled by the Langevin equation
\begin{align}\label{full_langevin}
\dot{x}(t) = v(t)\,p(t)  + \gamma^{-1}F + \eta(t),
\end{align}
where $\gamma$ is the friction coefficient and the force on the particle is generated from the external potential $\phi(x)$ according to $F\!=\!-\partial \phi/\partial x$. The distance is measured in units of $d$, the particle diameter and the time is measured in units of $d^2/D_t$, where $D_t$ is the translational diffusion constant. The stochastic term $\eta(t)$ is Gaussian distributed with zero mean and has time correlation $\langle\eta(t)\eta(t')\rangle=2D_t\delta(t-t')$. As in Ref. \cite{sharma2017escape}, we choose the following form for the external potential:

\begin{equation}\label{potential}
\beta \phi(x)=\frac{1}{2}\omega_0 x^2 - \alpha |x|^3,
\end{equation}
where $\beta = 1/k_{\rm B}T$, and $\omega_0$ and $\alpha$ are parameters (see Fig. \ref{sketch}). We note that the chosen potential is not special, and the presented approach is valid for a wide range of potentials. The third derivative is non-analytic at the origin, however, this feature does not influence the results presented below.

An intuitive choice for the time evolution of the propulsion speed is a decreasing function. We propose a class of models that can be used to describe such decay, the Generalised Exponential Model of order $n$ (GEM-$n$). We assume the self-propulsion speed to have the specific form:
\begin{equation}\label{self_propulsion}
v(t)=v_0\ e^{-\left(t/t_0\right)^n},
\end{equation}
where $v_0$ is the initial self-propulsion speed and $t_0$ is the characteristic decay time. We note that the specific choice of functional form used to model the decay of the activity is not important. The exponent $n$ controls the rate of transition from active to passive state; for $n=0$ the particle is always active and for $n\rightarrow \infty$ the activity drops to zero discontinuously at $t = t_0$.

We are interested in the escape rate of a particle starting at the origin that escapes over the potential barrier and is captured by a sink located sufficiently far from the barrier. It can be shown, in the case of constant self-propulsion speed $v(t)=v_0$, that in the long time limit $t \gg \tau$ the probability density distribution of a freely diffusing ABP (no external potential) reduces to a Gaussian with the diffusion constant \cite{balakrishnan2005connection, scacchi_sharma}
\begin{equation}\label{effective_diffusion_coefficient}
D =  D_t + D_a=D_t+\frac{v_0^2 \tau}{2},
\end{equation}
where $D_a = v_0^2 \tau /2$ is the coefficient of diffusion due to the active motion of the particle. 

The effective equilibrium approach~\cite{farage2015effective,sharma2017escape} describes ABPs in an external potential. In this approach, one obtains an approximate Fokker-Planck equation with an effective external potential $\phi^{\rm eff}(x)$ and an effective position-dependent diffusion constant $D(x)$:
\begin{align}
D(x) &= D_t+\frac{D_a}{1+\frac{\tau D_t}{d^2} \beta \phi^{\prime\prime}(x)}\, \label{Dx},\\
\beta \phi^\text{eff}(x)&=\int_0^x dy\,\frac{\beta \phi'(y)+D'(y)/D_t}{D(y)/D_t} \label{Vx}.
\end{align}
The effective diffusion constant is determined jointly by the activity and the external potential. This last quantity will reduce to the expression in Eq.~(\ref{effective_diffusion_coefficient}) when there is no external potential, corresponding to free diffusion.

The expressions in Eqs.~(\ref{Dx}) and~(\ref{Vx}) were obtained under the assumption~\cite{farage2015effective,sharma2017escape,wittmann2017effective} that the stochastic process corresponding to time evolution of the orientation vector can be considered as a Gaussian noise process with a finite correlation time. However, the mapping is only approximate because the process is not Gaussian. This is most evident in one dimension where the process is a random telegraphic process. However, as shown in Ref. \cite{scacchi_sharma}, the effective equilibrium approach provides an excellent description of an escaping ABP in one dimension. As we show below the approach also yields accurate description in the case of transient activity. For the sake of completeness, the expression for the escape rate, $r_{act}$, for a constant self-propelled velocity, obtained in the effective equilibrium approach is given as~\cite{sharma2017escape}
 \begin{align}\label{MFPT_EFF_EQ}
r_{act} =r_\text{pass}\exp{\left({\frac{D_a\left(\beta E_0 - \omega_0  \tau D_t/d^2 \right)}{D_t+D_a}}\right)},
 \end{align}
 where $r_{\rm pass} = \beta D_t\omega_0\exp(-\beta E_0)/(2\pi)$ is the escape rate of a passive particle over a single potential barrier described by Eq.~\eqref{potential} and $\beta E_0 = \omega_0^3/(54\alpha^2)$ is the height of the barrier. 

In case of transient activity as in Eq. \eqref{self_propulsion}, $n\rightarrow \infty$ corresponds to the case of a Brownian particle with constant activity $v_0$ for $0\leq t \leq t_0$. For $t>t_0$, the particle is passive. We first calculate the mean escape time for this particular case in an approximate fashion. We numerically obtain the distribution of the first passage time $\xi$ for an ABP with constant activity. We then numerically integrate this distribution over the time interval $0\leq t \leq t_0$. This integral, which we refer to as $\varrho$, is the fraction of the total number of particles which escape the potential barrier before $t=t_0$, i.e., the particle escapes the barrier in an active state. We assume that the rest of the particles, with the fraction $(1-\varrho)$, escape the barrier in passive state. We then approximate the mean escape rate $r$ of the transiently active Brownian particle as a weighted sum of these two contributions:
\begin{equation}\label{rate}
r =\left(\frac{\varrho}{r_{act}}  + \frac{1-\varrho}{r_{pass}}\right)^{-1},
\end{equation}
where $r_{act}$ and $r_{pass}$ are the mean escape rates for active and passive particles, respectively (Eq. \eqref{MFPT_EFF_EQ}). For more accurate results $r_{act}$ can be replaced by the average escape rate extracted from the distributions of the simulated first passage times $\xi$ for an ABP with constant activity. In the second term of the above expression, the position of the (passive) particle at $t=t_0$ is completely ignored. It is assumed that the mean escape time of the particle is the same as that of a passive particle starting at origin. This is a rather strong assumption. Even though the particle ceases to be active after $t_0$ it would escape the barrier faster than a passive particle starting at the origin. Therefore, the expression in Eq. \eqref{rate} is an underestimation of the escape rate. 

Whereas the time scale $t_0$ clearly marks the onset of passive behaviour in the case of $n\rightarrow \infty$, there is no such distinction (for $t \geq t_0$) between active and passive state for finite $n$. Despite this limitation, we continue to use the rate expression in Eq. \eqref{rate} for finite $n$. For finite $n$, we redefine a characteristic time scale of activity as 

\begin{equation}\label{effective}
\tilde{t}_0=\frac{t_0}{2^{1/n}}.
\end{equation}
$\tilde{t}_0$ is the time scale at which the active diffusion coefficient reduces by a factor of {\it e}. Admittedly, this rescaled time scale is defined in an arbitrary fashion and is model dependent. Nevertheless, it allows us to test the usefulness of Eq. \eqref{rate} in predicting the escape rate of a transiently active Brownian particle for all $n$. With this new time scale, for finite $n$, the fraction of particles escaping over the barrier in active state is estimated as the integral of the first passage time distribution over the time interval $0\leq t \leq \tilde{t}_0$.

Before presenting the results, we give a brief description of the simulations performed in this study. We set the particle size to $d=1$. Time is always measured in units of $d^2/D_t$ and since we have chosen $D_t = 1$ together with $k_{\rm B}T = 1$, the friction coefficient takes the value $\gamma = 1$. Eq.~\eqref{full_langevin} is integrated in time to generate the particle trajectory by advancing time in steps of $dt=2\cdot 10^{-3}$. The integral of the stochastic process $\eta(t)$ over a time interval $dt$ is taken as a Gaussian distribution with zero mean and variance $\int_0^{dt}\int_0^{dt}\langle\eta(s)\eta(s')\rangle ds ds' = 2D_t dt$. At every time step, the orientation of the particle is flipped with a probability $dt/\tau$. Without loss of generality, the persistence time is fixed for the remainder of this work at a value $\tau=5\cdot10^{-2}$. 

\section{Results and discussion}
\begin{figure}[h!]
\includegraphics[width=0.5\textwidth]{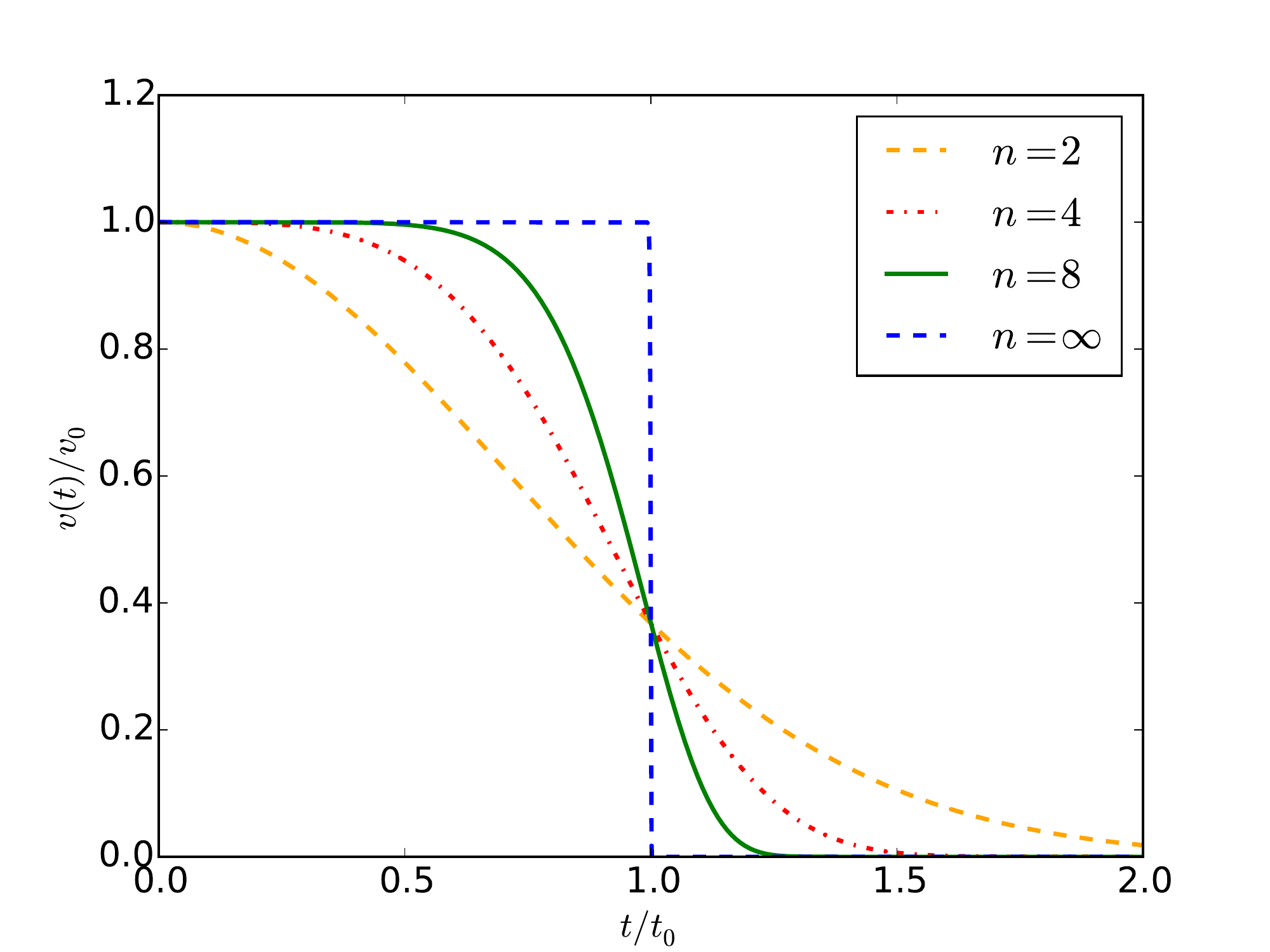}
\caption{Different self-propelled speed functions for varying exponent $n$.}\label{activity_dependence_plot}
\end{figure}
\begin{figure}[h]
    \centering
    \begin{tikzpicture}
        \node[anchor=south west,inner sep=0] (image) at (0,0) {\includegraphics[width=9.4cm]{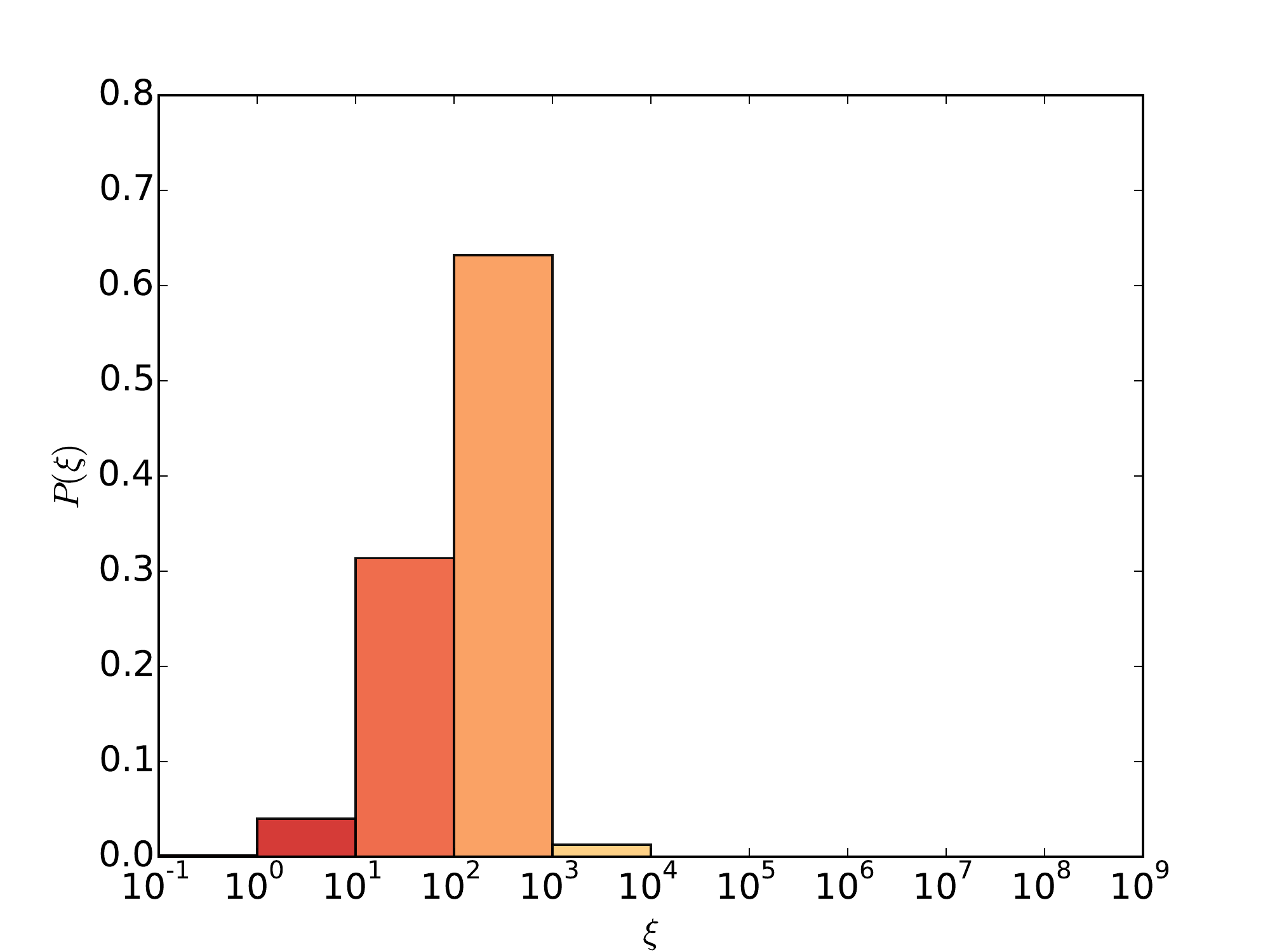}};
        \begin{scope}[x={(image.south east)},y={(image.north west)}]
            \node[anchor=south west,inner sep=0] (image) at (0.436,0.436) {\includegraphics[width=0.241\textwidth]{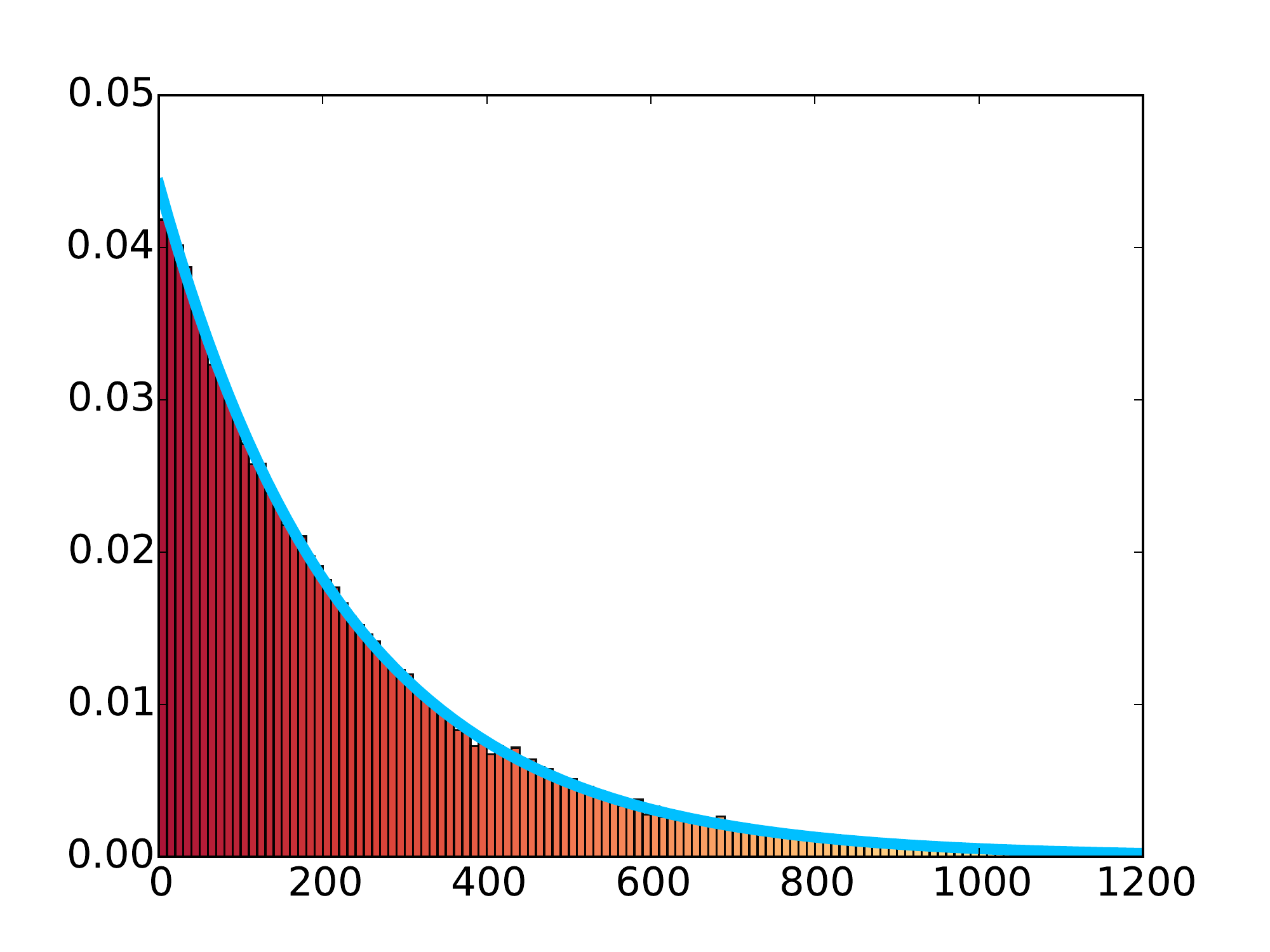}};
        \end{scope}
    \end{tikzpicture}
    \caption{Probability distribution of the FPT for active particles with constant activity. In the inset, the same quantities on a linear scale. The data is fitted (solid line) by a Gamma distibution of the form $(\Gamma(k) \theta)^{-1}t^{k-1}e^{-t/\theta}$, where $k$ is the shape parameter and $\theta$ is the scale parameter and $\Gamma$ the Gamma function. We obtained $k=1.00$ and $\theta=\theta^A\approx 228$ numerically. Number of simulated trajectories $N=75000$.}\label{active}
\end{figure}
\begin{figure}[h]
\includegraphics[width=0.5\textwidth]{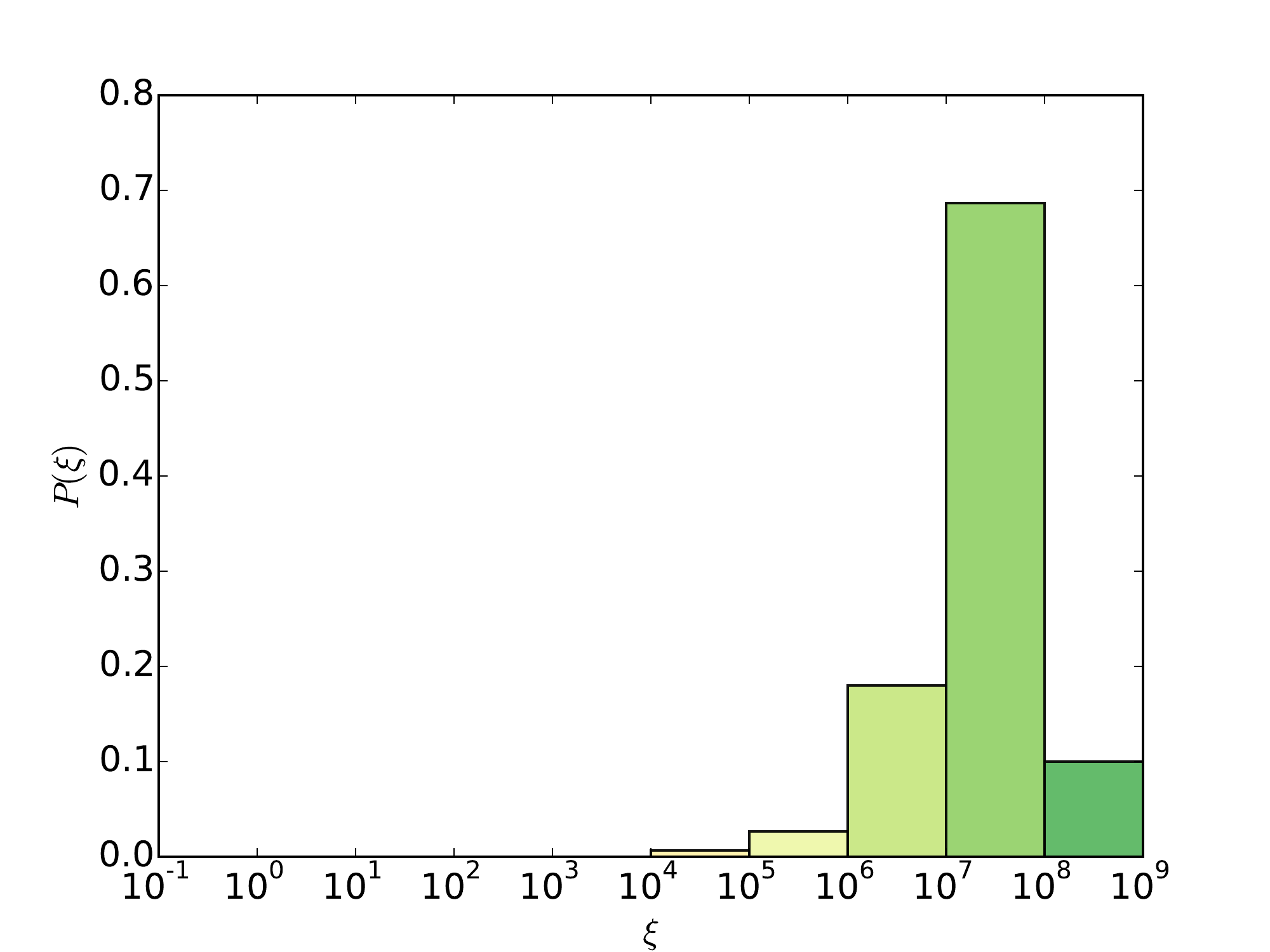}
\caption{Probability distribution of the FPT for passive particles. Number of simulated trajectories $N=300$.}\label{passive}
\end{figure}
\begin{figure}[h!]
\begin{tabular}{@{}c@{}}
\includegraphics[width=0.50\textwidth]{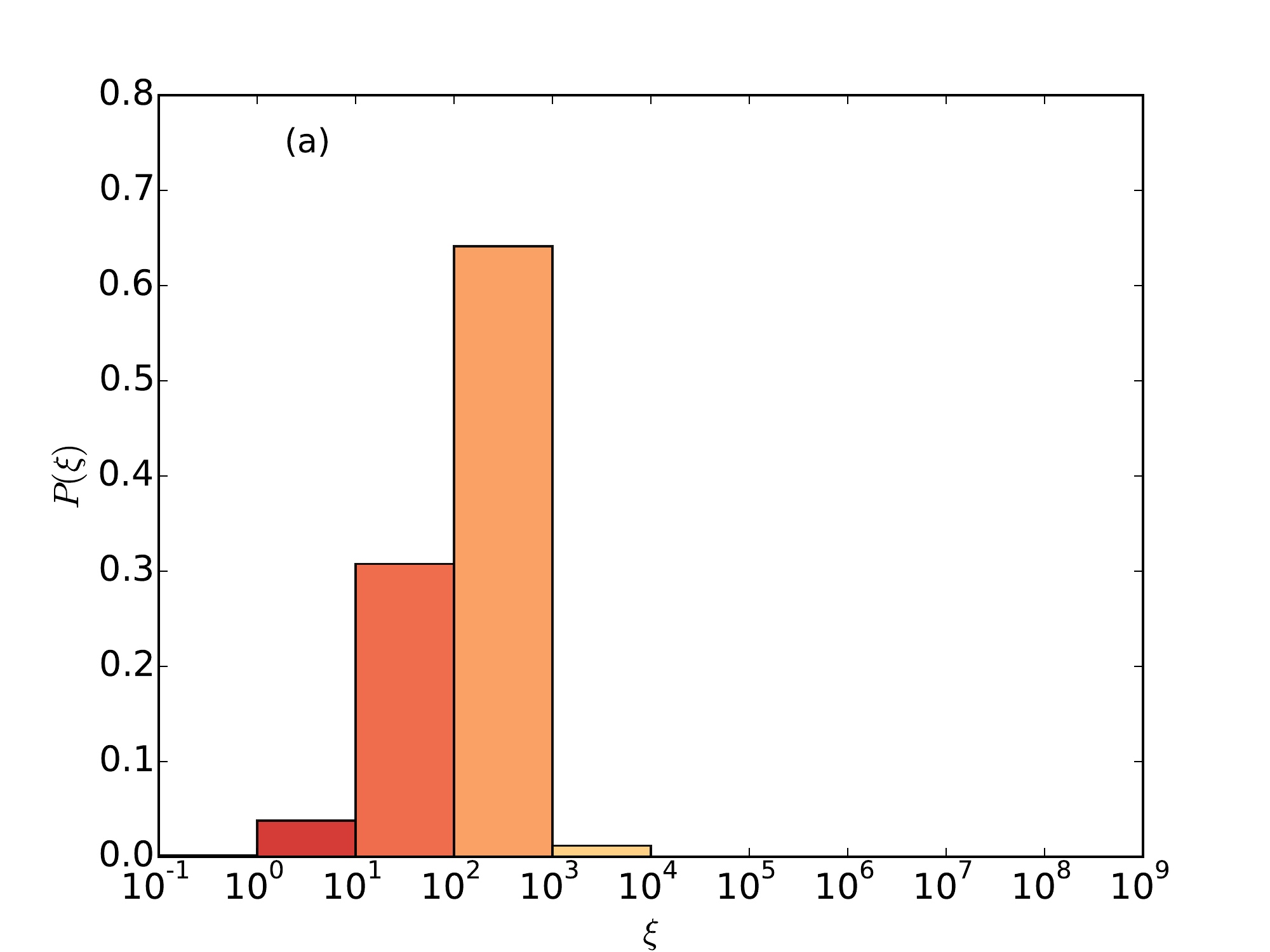}\\
\includegraphics[width=0.50\textwidth]{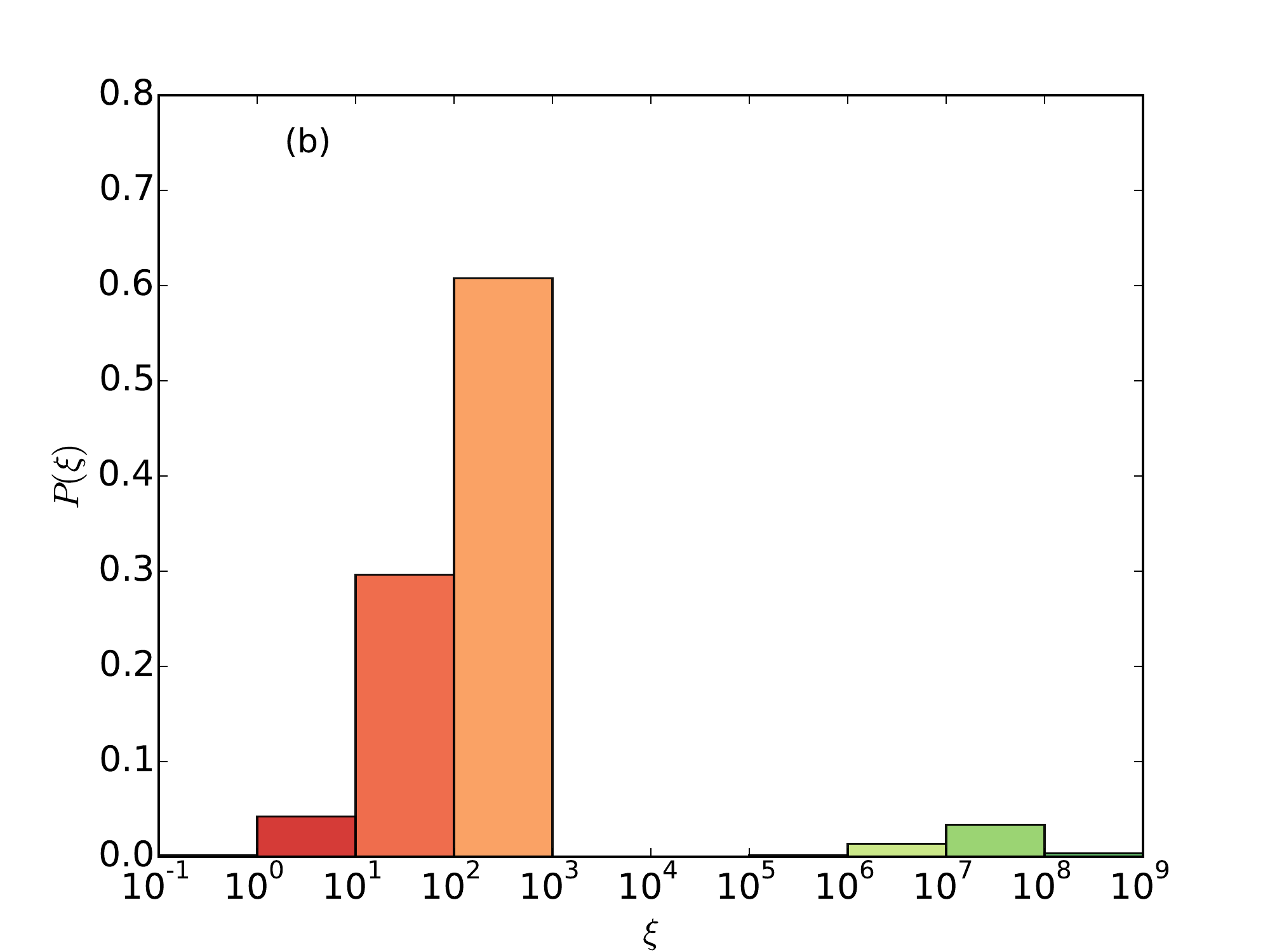}\\
\includegraphics[width=0.50\textwidth]{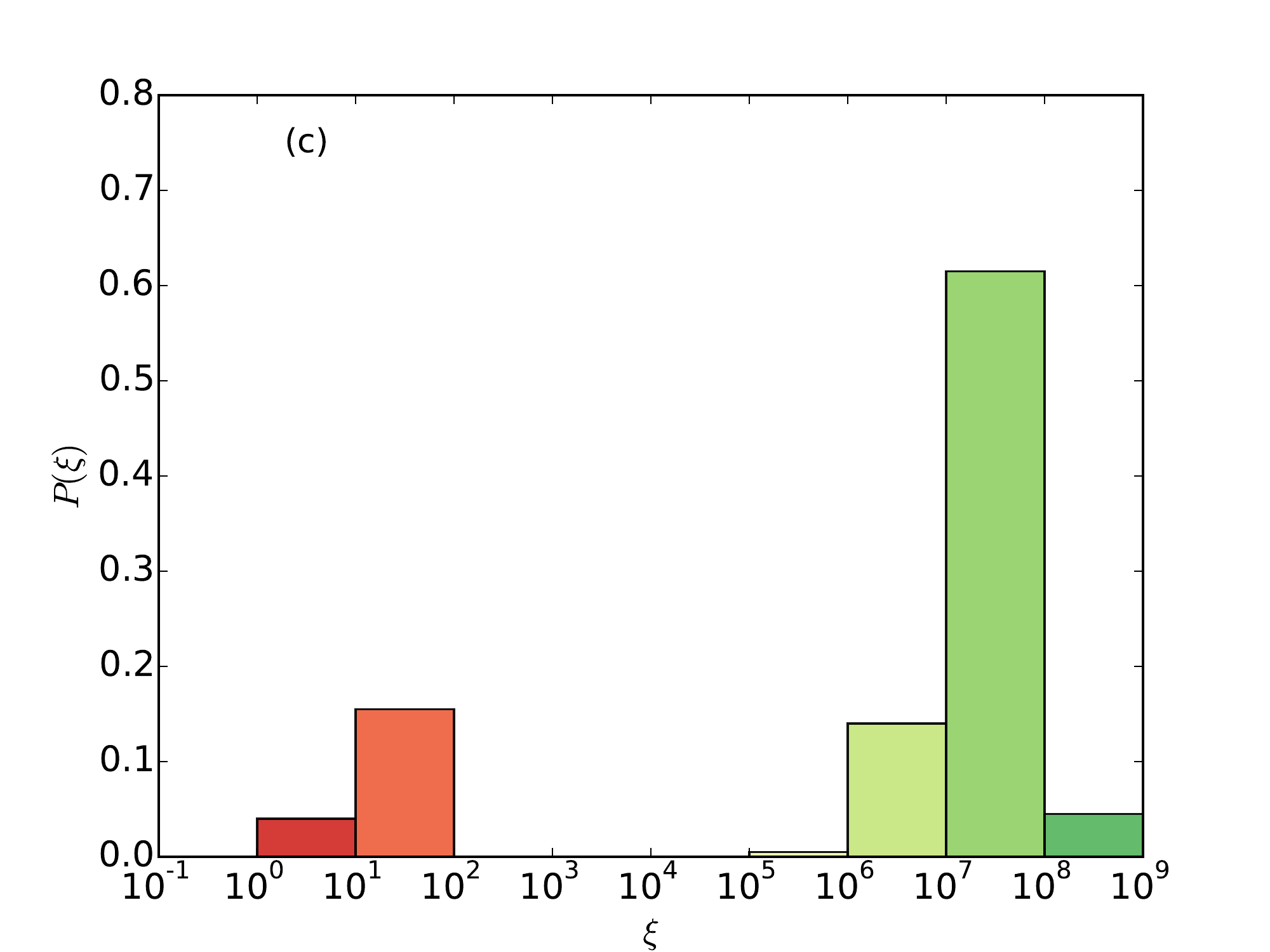}
\end{tabular}
\caption{In (a) $t_0=10^4$, $N=2\cdot10^4$. The probability distribution looks very close to the one obtained in the case of constant activity, meaning that most of the particles escaped the potential before loosing the activity. In (b) $t_0=10^3$, $N=800$. In (c) $t_0=100$, $N=200$.}
\label{diff_t_0_distributions}
\end{figure}
\begin{figure}[h!]
\includegraphics[width=0.52\textwidth]{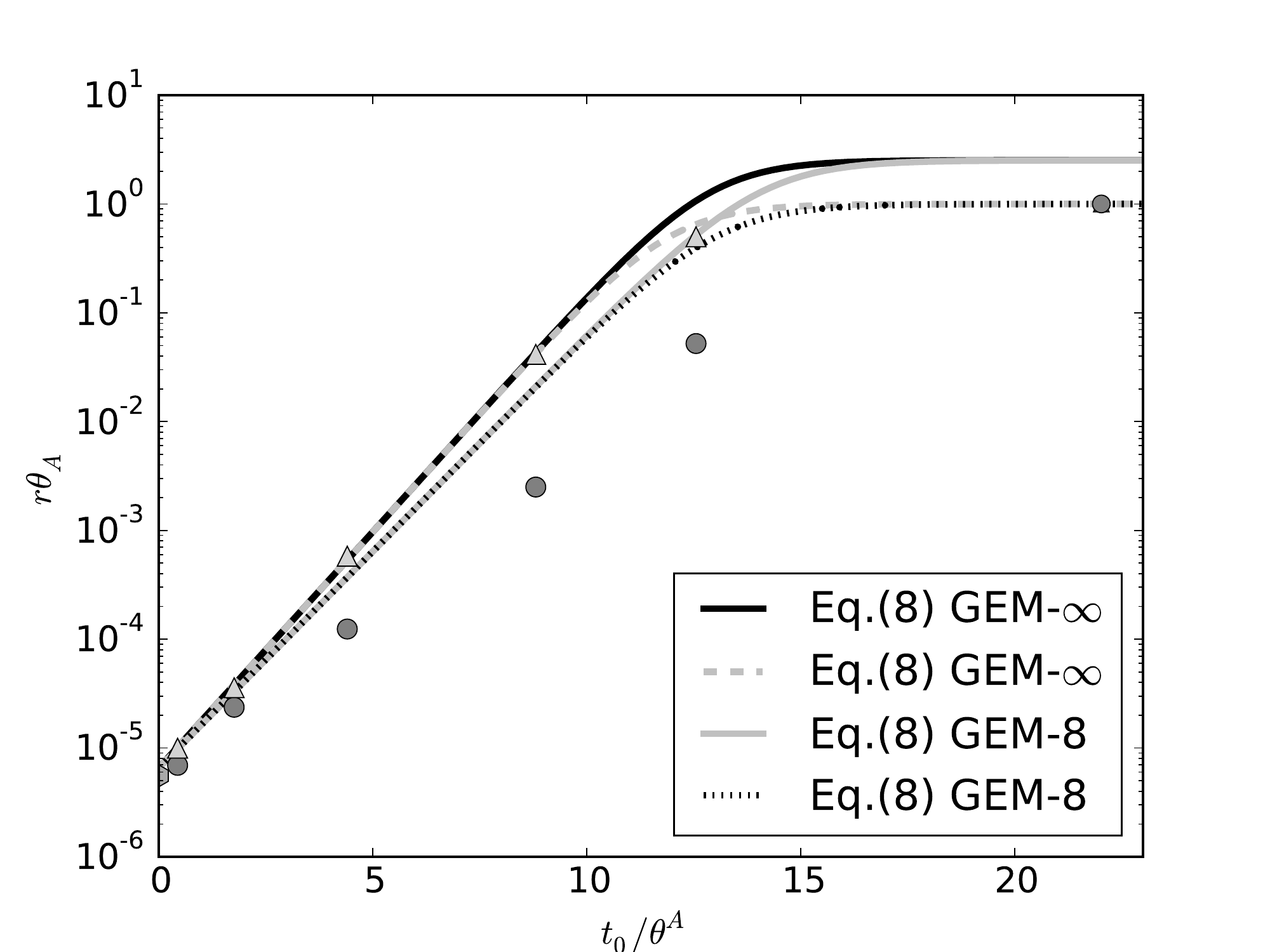}
\caption{GEM-$\infty$ particles: the dark solid line represents the prediction of Eq.~(\ref{rate}) where the value of $r_{act}$ is obtained from Eq.~(\ref{MFPT_EFF_EQ}) and the light dashed line represents the prediction of Eq.~(\ref{rate}) using $r_{act}=\theta_A^{-1}$. 
GEM-$8$ particles: the light solid line represents the prediction of Eq.~(\ref{rate}) where the value of $r_{act}$ is obtained from Eq.~(\ref{MFPT_EFF_EQ}) and the dark dotted line represents the prediction of Eq.~(\ref{rate}) using $r_{act}=\theta_A^{-1}$.
Numerical simulations: passive Brownian particles (hexagon), GEM-$8$ particles (circles) and GEM-$\infty$ particles (triangles).}
\label{theory_check}
\end{figure}
In Fig.~(\ref{activity_dependence_plot}) we show different self-propelled speed functions for varying exponent $n$. In the first part of this section we focus on the case where the exponent is set to $n=8$. The particles are trapped in an external potential well described by Eq.~(\ref{potential}). We discuss a specific set of parameters: $\alpha=1$ and $\omega_0=10$. These parameters are chosen to be the same as in Refs. \cite{sharma2017escape} and \cite{scacchi_sharma}. The initial self-propelled speed is set to $v_0=10$, leading to an initial rotational diffusion coefficient $D_a^0=2.5$. In Fig.~(\ref{active}) we show the simulation results in the case of constant activity, i.e. $t_0\rightarrow \infty$. As we can see, the first passage times (FPTs) $\xi$ are mostly concentrated on short time scales, between roughly $t=1$ and $t=10^4$. On the other hand, in the case of passive particles, i.e. $v_0=0$, the single events span over much longer time scales, between roughly $t=10^5$ and $t=10^9$ (see Fig.~(\ref{passive})). For these two extreme cases, the mean first passage time (MFPT) is well estimated by the theory presented in Ref.~\cite{scacchi_sharma} and can be obtained analytically using Eq.~(\ref{MFPT_EFF_EQ}). The rate of escape can be obtained as the inverse of the MFPT. We focus now on intermediate situation, where the activity decreases in time and behaves according to Eq.~(\ref{self_propulsion}). We compare hereafter these results with the distributions obtained in Figs.~(\ref{active}, \ref{passive}).
We denote the MFPT of the active particles with constant activity as $\theta^A$. For the set of parameters used here $\theta^A\approx 228$. In Fig.~(\ref{diff_t_0_distributions}.a) we show the results for $t_0=10^4$, corresponding to a ratio $t_0/\theta^A\approx 44$. As can be seen, the distribution is very similar to the one obtained in Fig.~(\ref{active}), which is an indication that the particles escaped the barrier before losing their activities. In Fig.~(\ref{diff_t_0_distributions}.b) we look at the case where $t_0=10^3$, corresponding to a ratio $t_0/\theta^A\approx 4.4$. It appears clear that some particles could not escape from the barrier before losing their activities. This fact will have a strong influence on the MFPT, as we will discuss later. Finally, in Fig.~(\ref{diff_t_0_distributions}.c) the results are for $t_0=100$, corresponding to a ratio $t_0/\theta^A\approx 0.44$. From this distribution we can conclude that the majority of the particles behave like passive particles. By decreasing the characteristic time $t_0$, the distribution will converge to the one corresponding to passive particles. Fig.~(\ref{diff_t_0_distributions}) confirms the hypothesis that the FPTs arise either from particles behaving actively or passively. As one can see in Figs.~(\ref{diff_t_0_distributions}.b, \ref{diff_t_0_distributions}.c), in those cases where the two times scales are on the same order, i.e. $t_0/\theta^{A}\approx1$, the MFPT is more complicated to treat. The two limiting cases (i) $t_0/\theta^A\gg 1$ and (ii) $t_0\ll \theta^{A}$ can be understood as follows. In the case described by (i), the MFPT of a particle with time dependent activity can be approximated by $\theta^A$. In the situation described in (ii), the MFPT can be approximated by the MFPT of passive particles $\theta^{P}$. To estimate the MFPT for the intermediate cases we use Eq.~(\ref{rate}). In Fig.~(\ref{theory_check}) we compare our predictions with numerical simulations [note that Eq.(\ref{rate}) describes the rate of escape over a single barrier, however the simulations are performed in a potential well. One has to divide the numerically obtained rates by a factor of 2]. As one can see, both the GEM-$8$ and the GEM-$\infty$ models are consistent with our simple theory. The simulated distribution of the first passage time of an active particle is well fitted by a Gamma distribution (see inset Fig.~(\ref{active})). The wieght $\varrho$ can therefore be expressed as $\varrho=\int_{0}^{t_0}\frac{1}{\theta^{A}}e^{-t/\theta^A}dt$. Using the latter form in Eq.~(\ref{rate}) the rate of escape increases exponentially for small $t_0/\theta_A$. This behaviour is evident in Fig.~(\ref{theory_check}).

\section{Conclusion}
We investigated the mean first passage time of an Active Brownian Particle with time dependent activity escaping a barrier in one dimension using numerical simulations. The particle undergoes a telegraphic process described by the average rate $\tau$ and with a self-propulsion speed decreasing in time. We have modelled the time dependence of the activity with a step function function taking either the maximum self-propelled speed $v_0$ or zero (passive particle). This allows us to estimate the mean first passage time by simply performing a weighted sum of contributions arising either from a constantly active particle or a passive particle. In order to perform analytics, we made the assumption that the motion of a freely diffusing ABP can be described at all time by the ordinary diffusion equation with the diffusion constant given in Eq.~\eqref{effective_diffusion_coefficient}. This is a reasonable assumption only under the self consistency condition that the MFPT is much larger than $\tau$, which is a limitation of this approach. The theoretical predictions of Eq.~(\ref{rate}) are in good agreement with our numerical simulations. We chose to describe the activity with the so called Generalised Exponential Model (GEM) of order $n=\infty$ and $n=8$, however the approach that we propose can be applied to different time dependent activities, as long as the two MFPTs: of the active particles with constant activity $\theta^A$ and the MFPT of the passive particles $\theta^P$, are widely separated from each other, that is $\frac{\theta^P}{\theta^A}\gg 1$, and the transition from active particle to passive particle happens on a short time scale. The validity of the theoretical prediction is limited to low/medium activities. We focussed on a single trapping potential, but this approach can be extended to any other external potential. In this work we focussed on a simple 1-dimensional problem. It will be of interesting to extend this approach to higher dimensions  with the goal of treating more realistic problems. The richer case of gradually changing activity will be addressed in the near future.

\section{Acknowledgements}
Alberto Scacchi thanks the Swiss National Science Foundation for financial
support under the grant number P2FRP2$\_$181453 and Tapio Ala-Nissila, together with Andrew Archer, for technical support.

\bibliographystyle{apsrev}
\bibliography{references}

\end{document}